\documentclass[11pt,a4paper]{article}
\usepackage{amssymb,amsmath,amsthm}
\usepackage{hyperref}
\usepackage[utf8]{inputenc}

\newcommand{\Integer}{\mathbb{Z}}
\newcommand{\Real}{\mathbb{R}}
\newcommand{\Complex}{\mathbb{C}}

\newcommand{\mat}{\begin{pmatrix}}
\newcommand{\tam}{\end{pmatrix}}

\newcommand{\idop}{\mathbb{I}}

\newcommand{\g}{\mathfrak{g}}
\newcommand{\h}{\mathfrak{h}}  
\newcommand{\ag}{\mathfrak{\hat{g}}}
\newcommand{\ah}{\mathfrak{\hat{h}}}  
\newcommand{\fI}{\mathfrak{I}}  
\newcommand{\fT}{\mathfrak{T}}

\DeclareMathOperator{\sdim}{\text{sdim}}
\DeclareMathOperator{\str}{\text{str}}

\DeclareMathOperator{\ad}{{\text{ad}}}

\title{On conformal field theories based on Takiff superalgebras}
  
\author{Thomas Quella\\[2mm]\\The University of Melbourne\\School of Mathematics and Statistics\\Parkville 3010 VIC, Australia
\\[5mm]{\sc Thomas.Quella@unimelb.edu.au}}

\date{}

\begin{document}
\maketitle
\begin{abstract}
  We revisit the construction of conformal field theories based on
  Takiff algebras and superalgebras that was introduced by Babichenko and
  Ridout. Takiff superalgebras can be thought of as truncated current
  superalgebras with $\Integer$-grading which arise from taking $p$ copies of a Lie
  superalgebra $\g$ and placing them in the degrees
  $s=0,\ldots,p-1$. Using suitably defined non-degenerate invariant
  forms we show that Takiff superalgebras give rise to families of
  conformal field theories with central charge $c=p\sdim\g$. The
  resulting conformal field theories are defined in the standard way,
  i.e.\ they lend themselves to a Lagrangian description in terms of a WZW model and
  their chiral energy momentum tensor is the one obtained naturally
  from the usual Sugawara construction. In view of their intricate
  representation theory
  they provide interesting examples of conformal field theories.
\end{abstract}

\section{Introduction}

  Conformal field theories (CFTs) based on affine Lie algebras and superalgebras are the basic building block for many constructions within CFT, for instance the GKO coset construction \cite{Goddard:1985vk}, orbifolds \cite{Felder:1988CMaPh.117..127F} or quantum Hamiltonian reduction \cite{Bershadsky:1989mf,Feigin:1990pn}.
  Also known under the name Wess-Zumino-Witten models (WZW models), the defining data of such CFTs consists of a finite-dimensional Lie (super)algebra $\g$, an associated Lie (super)group $G$ and an associated invariant bilinear form $\langle\cdot,\cdot\rangle:\g\otimes\g\to\Complex$ with suitable non-degeneracy properties.
  The basic properties of WZW models have been illucidated in a series of papers \cite{Witten:1983ar,Knizhnik:1984nr,Gepner:1986wi}, both from a geometric and an algebraic perspective.

  Historically, the attention mainly focused on WZW models based on simple or abelian Lie algebras. With these ingredients one can readily understand WZW models based on (compact) reductive groups. The first example of a WZW  model based on a non-reductive group is the famous Nappi-Witten model \cite{Nappi:1993ie} which describes string theory on a plane wave background. This paper immediately triggered a lot of activity in this area which is neatly summarized in \cite{FigueroaO'Farrill:1994hx,Figueroa-OFarrill:1996NuPhB.458..137F}. One of the milestones was the exact and complete solution of the $H_4$ plane wave model \cite{D'Appollonio:2003dr,D'Appollonio:2003ym,Bianchi:2004vf,D'Appollonio:2004pm} which is based on the \.Inönü-Wigner contraction or Penrose limit of $SL(2,\Real)\times U(1)$.

  More recently, Babichenko and Ridout took up the subject again in an effort to provide new examples of solvable logarithmic conformal field theories \cite{Babichenko:2013JPhA...46l5204B}. They introduced a special class of CFTs based on affine Takiff superalgebras which were obtained by combining $p=2$ copies of the underlying current algebra in an indecomposable way. In a different line of research, Rasmussen and Raymond studied what they called Galilean contractions of affine Lie algebras and W-algebras \cite{Rasmussen:2017NuPhB.922..435R,Rasmussen:2019NuPhB.94514680R}, thereby arriving at order $p$ generalizations of affine Takiff algebras. It should be noted that both of these constructions were completely algebraic and in some regards seemed rather ad hoc, at least from a physical perspective.

  The goal of the present note is to put the results of   \cite{Babichenko:2013JPhA...46l5204B,Rasmussen:2017NuPhB.922..435R,Rasmussen:2019NuPhB.94514680R} in an appropriate structural context and point out that the construction is indeed very natural, from both a physical and mathematical perspective. In particular, we emphasize the point that the conformal field theories considered in \cite{Babichenko:2013JPhA...46l5204B} can be regarded as arising from genuine WZW models. Together with general results on Lie (super)algebras with non-degenerate invariant form \cite{FigueroaO'Farrill:1994hx,Figueroa-OFarrill:1996NuPhB.458..137F} this implies the existence of a conformal energy-momentum tensor which has central charge $c=p\sdim(\g)$. Maybe the most important aspect of our work is that it enables the use of geometric methods in the solution of these models. For the special case of a Takiff superalgebra based on $GL(1|1)$ this has already been employed in \cite{Babichenko:2015SIGMA..11..067B}.

  The paper is organized as follows. In Section~\ref{sc:FiniteTakiff} we first of all introduce Takiff superalgebras of arbitrary order $p$ and discuss some of the associated structures such as invariant forms and automorphisms. The affinization of general finite dimensional Lie superalgebras with invariant form is reviewed in Section~\ref{sc:Affinization}. It is shown that the affinization of Takiff superagebras gives rise to a conformal energy momentum tensor by means of the Sugawara construction and hence to a CFT. We also argue that (or rather in which sense) this construction gives the same result as applying the Takiff construction to affine Lie superalgebras. Finally, Section~\ref{sc:Examples} applies the abstract considerations to specific examples in order to relate this work to the existing literature before the Conclusions summarize the present work and point out directions for future research.

\section{\label{sc:FiniteTakiff}Finite dimensional Takiff
  superalgebras}

\subsection{\label{sc:FiniteTakiffDefinition}Definition and structure}

  The basic theory of Lie superalgebras was developed by Kac \cite{Kac:1977em}. For our current paper we only need two relatively simple concepts, the definition of a Lie superalgebra and the notion of a metric.\footnote{None of the statements below requires a non-trivial odd part $\g_{\underline{1}}$. Our analysis thus equally well applies to Takiff {\em algebras}.} A {\em Lie superalgebra} $\g$ is a $\Integer_2$-graded vector space $\g=\g_{\underline{0}}\oplus\g_{\underline{1}}$ with a Lie bracket $[\cdot,\cdot]:\g\otimes\g\to\g$ which respects the grading in the sense that $[\g_i,\g_j]\subset\g_{i+j}$ and, moreover, satisfies the usual properties such as graded anti-symmetry and graded Jacobi identity \cite{Kac:1977em}. A {\em metric} is an even bilinear form $\langle\cdot,\cdot\rangle:\g\otimes\g\to\Complex$ which is non-degenerate, graded symmetric and which, in addition, satisfies the invariance property
\begin{align}
  \label{eq:Invariance}
  \langle[X,Y],Z\rangle=\langle X,[Y,Z]\rangle\;.
\end{align}
  Throughout the text we will work with a basis of generators $J^a$ of $\g$. The generators will always be assumed to be
  homogenous elements of $\g$, i.e.\ have a well-defined degree $d_a\in\Integer_2$. With respect to this fixed basis we can define the structure constants ${f^{ab}}_c$ and metric tensor $\kappa^{ab}$ via
\begin{align}
  [J^a,J^b]=i{f^{ab}}_cJ^c
  \qquad\text{ and }\qquad
  \kappa^{ab}=\langle J^a,J^b\rangle\;.
\end{align}
  The two tensors ${f^{ab}}_c$ and  $\kappa^{ab}$ satisfy obvious relations that reflect the structural properties of $\g$ and its metric that have been mentioned before.

  Following the reasoning of \cite{Babichenko:2013JPhA...46l5204B} (see also
  \cite{Takiff:MR0281839}) we would like to extend the algebra $\g$ by
  taking $p\geq2$ copies of it and by defining a new Lie bracket on
  the resulting space $\fT_p(\g)=\g^{(0)}\oplus\cdots\oplus\g^{(p-1)}$. Here $\g^{(s)}=\g$
  as vector spaces and the $\Integer_2$-grading is simply inherited from $\g$. The construction becomes most transparent if we introduce a
  formal nilpotent even variable $\Theta$ that satisfies $\Theta^p=0$ but
  $\Theta^{p-1}\neq0$. The space $\g^{(s)}$ can then be identified with
  the vector space $\g\otimes\Complex\Theta^s$ and the Lie bracket is
  defined by the simple assignment
\begin{align}
  \label{eq:TakiffDefinition}
  [X\otimes\Theta^r,Y\otimes\Theta^s]
  =[X,Y]\otimes\Theta^{r+s}\;,
\end{align}
  where $X,Y\in\g$.
  Using the properties of $\g$, it can easily be checked that this
  definition gives rise to a Lie superalgebra. Following the
  suggestion of \cite{Babichenko:2013JPhA...46l5204B}, we will call this Lie
  superalgebra $\fT_p(\g)$ a {\em Takiff superalgebra}.

  The Takiff superalgebra $\fT_p(\g)$ has a natural
  $\Integer$-grading which is localized in the degrees
  $s=0,\ldots,p-1$ and which has the original Lie superalgebra $\g$ as its grade $0$ Lie
  subsuperalgebra. By setting $\g^{(s)}=\{0\}$ for $s<0$ and $s\geq p$ we
  can write
\begin{align}
  \label{eq:Grading}
  \fT_p(\g)
  =\bigoplus_{s\in\Integer}\g^{(s)}\;.
\end{align}
  For any choice $s\in\Integer$ the space $\fI_s=\bigoplus_{r\geq s}\g^{(r)}$ is an ideal. In particular, we realize that the Lie subsuperalgebra $\fI_{\lfloor(p+1)/2\rfloor}$ is an abelian ideal, where the bracket denotes the integer part. The same is obviously true for all Lie subsuperalgebras of $\fI_{\lfloor(p+1)/2\rfloor}$, including the top degree subspace $\g^{(p-1)}$ of $\fT_p(\g)$. It is thus evident from the construction that $\fT_p(\g)$ is not semi-simple and generally not even reductive.
  
\subsection{\label{sc:InvariantForms}Invariant forms}

  The construction of the Takiff superalgebra $\fT_p(\g)$ suggests
  that all relevant structures such as invariant forms, automorphisms
  etc.\ are inherited from its grade $0$ Lie subsuperalgebra $\g$, at least
  as long as they are assumed to respect the grading. As we shall now
  discuss, this is indeed the case for the metric, at least if $\g$ is semi-simple. Let us start by defining natural
  metrics on $\fT_p(\g)$. For this purpose let
  $\langle\cdot,\cdot\rangle_s$ with $s=0,\ldots,p-1$ be a family of
  metrics on $\g$. This collection of metrics defines a metric on
  $\fT_p(\g)$ by using the assignment
\begin{align}
  \label{eq:InvariantForm}
  \langle X\otimes\Theta^r,Y\otimes\Theta^s\rangle
  =\langle X,Y\rangle_{r+s}\;.
\end{align}
  In this equation it is understood that the right hand side vanishes
  whenever $r+s\geq p$. It is easy to check that the form
  $\langle\cdot,\cdot\rangle$ defined in this way is even, graded
  anti-symmetric and invariant. Actually, in order to ensure the
  non-degeneracy of the metric it is only required that the top degree
  metric $\langle\cdot,\cdot\rangle_{p-1}$ is non-degenerate while the
  other metrics $\langle\cdot,\cdot\rangle_s$ with $s<p-1$ could equally
  well be degenerate.

  While it is straightforward to check that the assignment
  \eqref{eq:InvariantForm} satisfies all desired properties, it is
  important to know that all metrics have to be of this form, at least
  if $\g$ is semi-simple. In order to prove this assertion let us
  consider an arbitrary metric $\langle\cdot,\cdot\rangle$ on $\fT_p(\g)$. The first
  observation is that $\langle
  X\otimes\Theta^r,Y\otimes\Theta^s\rangle$ can only depend on $r+s$
  but not on $r$ and $s$ individually. Indeed, if $\g$ is semi-simple there
  exist elements $U_\alpha$ and $V_\alpha$ such that $X=\sum_\alpha[U_\alpha,V_\alpha]$. One then obtains the
  chain of equalities
\begin{align}
  \langle X\otimes\Theta^r,Y\otimes\Theta^s\rangle
  &=\langle\sum_\alpha[U_\alpha\otimes1,V_\alpha\otimes\Theta^r],Y\otimes\Theta^s\rangle
    =\sum_\alpha\langle U_\alpha\otimes1,[V_\alpha\otimes\Theta^r,Y\otimes\Theta^s]\rangle\nonumber\\[2mm]
  &=\sum_\alpha\langle U_\alpha\otimes1,[V_\alpha\otimes1,Y\otimes\Theta^{r+s}]\rangle
     =\sum_\alpha\langle[U_\alpha\otimes1,V_\alpha\otimes1],Y\otimes\Theta^{r+s}\rangle\nonumber\\[2mm]
  &=\langle X\otimes1,Y\otimes\Theta^{r+s}\rangle
     =:\langle X,Y\rangle_{r+s}\ .
\end{align}
  This proves that the metric has to be of the form   \eqref{eq:InvariantForm} but it leaves open whether the right hand   side is $\g$-invariant and graded symmetric. However, the latter properties follow easily from the corresponding properties of $\langle\cdot,\cdot\rangle$. Apart from proving our assertion, our calculation also shows that the metric has to vanish for $r+s\geq p$. If $\g$ fails to be semi-simple, there are more possibilities for defining metrics on $\fT_p(\g)$. This is obvious for Takiff superalgebras based on abelian Lie algebras $\g$ (where the $\Integer$-grading ceases to have a special meaning) and can also be verified in specific examples such as $\fT_2\bigl(gl(1|1)\bigr)$.

  Let us finally verify under which conditions the form defined in \eqref{eq:InvariantForm} is non-degenerate. For this purpose we fix an arbitrary element $X=\sum_{t=0}^{p-1}X_t\otimes\Theta^t\in\fT_p(\g)$ and assume the validity of the equation
\begin{align}
  0
  =\langle X,Y\rangle
  =\sum_{t=0}^{p-1}\langle X_t\otimes\Theta^t,Y\rangle
\end{align}
  for all choices of $Y\in\fT_p(\g)$. We then single out a specific $t_0\in\{0,1,\ldots,p-1\}$ and define $s_0=p-1-t_0$. The previous statement then implies that
\begin{align}
  0
  =\sum_{t=0}^{p-1}\langle X_t\otimes\Theta^t,Y_{s_0}\otimes\Theta^{s_0}\rangle
  =\langle X_{t_0},Y_{s_0}\rangle_{p-1}
\end{align}
  for all $Y_{s_0}\in\g$. If we assume the form $\langle\cdot,\cdot\rangle_{p-1}$ to be non-degenerate and iterate over all valid values of~$t$ it then follows that $X_t=0$ and hence $X=0$. It is important to emphasize that only the top form $\langle\cdot,\cdot\rangle_{p-1}$ needs to be non-degenerate while $\langle\cdot,\cdot\rangle_{r}$ for $0\leq r<p-1$ can well be degenerate. It is not possible to single out another component $r<p$ as a non-degenerate form instead since then $s_0=r-t_0$ would need to be negative for $t_0=p-1$.

  The invariant form in Eq.~\eqref{eq:InvariantForm} can be further simplified if we assume that $\g$ is simple. In this case all invariant forms are proportional to one standard non-degenerate form $K$ on $\g$ which for most of the cases can be chosen to be the Killing form.\footnote{Exceptions arise in parts of the $A$- and the $D$-series where the Killing form might vanish identically~\cite{Kac:1977em}.} In other words, we have $\langle\cdot,\cdot\rangle_r=k_r\,K(\cdot,\cdot)$. In our treatment of conformal field theories later on the constants of proportionality $k_r$ will play the role of what usually is called the level of a WZW model. A Takiff algebra $\fT_p(\g)$ based on a simple Lie superalgebra $\g$ thus admits $p$ distinct levels and non-degeneracy of the associated metric (only) requires $k_{p-1}\neq0$. In this special case we recover a relation to the description of affine Takiff superalgebras as higher order Galilean affine superalgebras as discussed in \cite{Rasmussen:2019NuPhB.94514680R}. Affine Takiff superalgebras will be discussed in more detail in Section~\ref{sc:Affinization}.

  It is well known that every Lie superalgebra admits a natural
  invariant form, the Killing form $K(\cdot,\cdot)$. For a Takiff superalgebra, the Killing form turns out to be generally degenerate even if the Killing form on $\g$ itself is non-degenerate. Indeed, a simple
  calculation yields
\begin{align}
  \label{eq:TakiffKilling}
  K(X\otimes\Theta^r,Y\otimes\Theta^s)
  =p\,\delta^{r0}\delta^{s0}\,K(X,Y)\ \ ,
\end{align}
  where on the right hand side the symbol $K(\cdot,\cdot)$ refers to
  the Killing form on $\g$. The previous equation can easily be
  obtained from the definition of the Killing form,
\begin{align}
  K(X\otimes\Theta^r,Y\otimes\Theta^s)
  =\str\bigl(\ad_{X\otimes\Theta^r}\circ\ad_{Y\otimes\Theta^s}\bigr)
  =\str\bigl(\bigl[X\otimes\Theta^r,[Y\otimes\Theta^s,\cdot]\bigr]\bigr)\ \ ,
\end{align}
  together with an explicit evaluation on a basis of generators,
\begin{align}
  \bigl[J^a\otimes\Theta^r,[J^b\otimes\Theta^s,J^c\otimes\Theta^t]\bigr]
  &=\bigl[J^a,[J^b,J^c]\bigr]\otimes\Theta^{r+s+t}\;.
\end{align}
  The factor $\delta^{r+s,0}=\delta^{r0}\delta^{s0}$ in
  \eqref{eq:TakiffKilling} arises due to the requirement $r+s+t=t$ that
  is imposed by taking the supertrace (recall that $r,s\geq0$). The factor $p$ results from the
  summation over $t=0,\ldots,p-1$. We wish to stress the fact that,
  even though the Killing form is degenerate (for $p\geq2$), the
  Takiff superalgebra $\fT_p(\g)$ admits whole families of natural
  non-degenerate invariant forms, see eq.\ \eqref{eq:InvariantForm}. Also, the Killing form \eqref{eq:TakiffKilling} is consistent with the form \eqref{eq:InvariantForm}, i.e.\ it may be thought of as being induced from a family of invariant forms on $\g$.
  
  In this paper we will exclusively be concerned with Takiff superalgebras $\fT_p(\g)$ which come equipped with the additional structure of an invariant and graded symmetric form. Except stated otherwise this invariant form will always be assumed to arise from a family of invariant forms on $\g$ in the sense of definition~\eqref{eq:InvariantForm}.
  
\subsection{\label{sc:Automorphisms}Automorphisms}

  For the study of representations of a Lie superalgebra but also for
  the investigation of D-branes in the associated WZW model it is
  important to have a detailed knowledge about its automorphisms, especially the isometric automorphisms.
  Let $\Omega$ be an automorphism of $\g$. The automorphism $\Omega$
  induces a grade-preserving automorphism on the associated Takiff
  superalgebra $\fT_p(\g)$ by setting
\begin{align}
  \Omega(J^a\otimes\Theta^s)
  =\Omega(J^a)\otimes\Theta^s\;.
\end{align}
  It can easily be checked by explicit calculation that this indeed
  defines an automorphism,
\begin{align}
  \bigl[\Omega(J^a\otimes\Theta^r),\Omega(J^b\otimes\Theta^s)\bigr]
  &=\bigl[\Omega(J^a)\otimes\Theta^r,\Omega(J^b)\otimes\Theta^s\bigr]
   =\bigl[\Omega(J^a),\Omega(J^b)\bigr]\otimes\Theta^{r+s}\\[2mm]
  &=\Omega\bigl([J^a,J^b]\bigr)\otimes\Theta^{r+s}
   =\Omega\bigl([J^a,J^b]\otimes\Theta^{r+s}\bigr)
   =\Omega\bigl([J^a\otimes\Theta^r,J^b\otimes\Theta^s]\bigr)\;.\nonumber
\end{align}
  If $\g$ comes equipped with a family of invariant forms $\langle\cdot,\cdot\rangle_s$ and $\Omega$ is isometric on $\g$ for {\em all} of these, the same will be true for the induced automorphism on $\fT_p(\g)$ with the natural induced metric. Indeed, a straightforward calculation gives
\begin{align}
  \bigl\langle\Omega(J^a\otimes\Theta^r),\Omega(J^b\otimes\Theta^s)\bigr\rangle
  &=\bigl\langle\Omega(J^a)\otimes\Theta^r,\Omega(J^b)\otimes\Theta^s\bigr\rangle
  =\bigl\langle\Omega(J^a),\Omega(J^b)\bigr\rangle_{r+s}\nonumber\\[2mm]
  &=\bigl\langle J^a,J^b\bigr\rangle_{r+s}
  =\bigl\langle J^a\otimes\Theta^r,J^b\otimes\Theta^s\bigr\rangle\;.
\end{align}
  This statement will turn out to be important in the context of studying affinizations in Section~\ref{sc:Affinization}.
  As a typical example let us mention the case of a simple Lie superalgebra $\g$. As pointed out in Section~\ref{sc:InvariantForms} in this case all invariant forms  $\langle\cdot,\cdot\rangle_s$ are proportional to a single invariant form and the condition on the isometry of $\Omega$ is much less restrictive than in the general case.

\section{\label{sc:Affinization}The affinization of finite dimensional
  Takiff superalgebras}

\subsection{\label{sc:AffinizationDefinition}Definition and Sugawara
  construction}

  In their paper \cite{Babichenko:2013JPhA...46l5204B}, Babichenko and Ridout
  showed that the application of the Takiff construction (with $p=2$)
  to an affine Lie superalgebra leads to a current superalgebra which
  gives rise to a conformal energy momentum tensor. A similar philosophy was adopted in \cite{Rasmussen:2017NuPhB.922..435R,Rasmussen:2019NuPhB.94514680R}. In fact, as we
  will point out now, it seems more natural to start with a finite
  dimensional Takiff superalgebra with a metric and then to define
  the associated affinization. Both constructions commute in a sense as will be shown,
  and should therefore be regarded as equivalent. However, our way of
  thinking allows for a geometric interpretation and establishes immediately that the conformal field theories
  considered in \cite{Babichenko:2013JPhA...46l5204B} are just ordinary WZW models
  (even though associated with non-reductive Lie groups). Using the knowledge that has
  been gained about WZW models on simple or reductive Lie groups
  \cite{Witten:1983ar,Knizhnik:1984nr,Gepner:1986wi} and, more
  recently, supergroups
  \cite{Schomerus:2005bf,Gotz:2006qp,Saleur:2006tf,Quella:2007hr,Quella:2013JPhA...46W4010Q}, our work opens
  a realistic perspective of being able to solve explicit examples, see also \cite{Babichenko:2015SIGMA..11..067B}.

  The affinization of a Lie superalgebra $\h$ with invariant form $\langle\cdot,\cdot\rangle$ follows the standard recipe. In order to simplify notation we will explain it for a general Lie superalgebra $\h$ with generators $J^A$.\footnote{The Takiff superalgebra $\fT_p(\g)$ is of course contained as a special case. However, it is not advisable to write down the expressions in the basis $J^a\otimes\Theta^s$ since there would be too many indices floating around. One can think of the index $A$ as a multi-label $(a,s)$.} The affinization $\ah$ of $\h$ (associated with the invariant form $\langle\cdot,\cdot\rangle$) is the central extension of the loop superalgebra $\h\otimes\Complex[t,t^{-1}]$ which is defined by the relations
\begin{align}
  \label{eq:Affinization}
  [X\otimes t^m,Y\otimes t^n]
  =[X,Y]\otimes t^{m+n}+m\,\langle X,Y\rangle\,\delta_{m+n,0}\,K\;,
\end{align}
  where $K$ denotes the central element. In physical applications, this algebra is always realized on representations where $K$ assumes a fixed number $k$. If $\h$ is simple and the invariant form is normalized appropriately $k$ is known as the level. In any case, after replacing $K$ by a number we can drop $K$ from Eq.~\eqref{eq:Affinization} and absorb the corresponding constant into a redefinition of the invariant form $\langle\cdot,\cdot\rangle$ and this convention will be understood from now on.
  
  The relation between affinizations and conformal field theories is best understood in terms of the Sugawara construction which assigns a Virasoro algebra to the affinization, at least for generic choices of the invariant form.\footnote{One will need to avoid the ``critical level'' and what this means will become clear below.} The Sugawara construction is best explained in terms of the currents $J^A(z)=\sum_{n\in\Integer}J_n^Az^{-n-1}$ which can be interpreted as the generating function for the modes $J_n^A=J^A\otimes t^n$.
  With the help of the matrix $\kappa^{AB}=\langle J^A,J^B\rangle$ and
  the structure constants $[J^A,J^B]=i{f^{AB}}_C\,J^C$ one can then rewrite
  the relation~\eqref{eq:Affinization} in the formalism of operator product expansions as
\begin{align}
  \label{eq:CurrentAlgebra}
  J^A(z)\,J^B(w)
  \sim\frac{\kappa^{AB}}{(z-w)^2}+\frac{i{f^{AB}}_C\,J^C(w)}{z-w}\;.
\end{align}
  It is well-known that the current algebra \eqref{eq:CurrentAlgebra}, for generic choices of $\kappa^{AB}$,
  implies the existence of a conformal energy momentum tensor $T(z)$
  which satisfies
\begin{align}
  \label{eq:Sugawara}
  T(z)\,T(w)
  &\sim\frac{c/2}{(z-w)^4}+\frac{2\,T(w)}{(z-w)^2}+\frac{\partial T(w)}{z-w}\\[2mm]
  T(z)\,J^A(w)
  &\sim\frac{J^A}{(z-w)^2}+\frac{\partial J^A(w)}{z-w}\;.
\end{align}
  The central charge $c$ is a characteristic of the underlying
  conformal field theory which depends on the Lie algebra $\h$ and on
  the metric $\langle\cdot,\cdot\rangle$.\footnote{Recall that we
    absorbed the usual level into the definition of the metric.} 
  In order to define $T(z)$ in terms of the currents $J^A(z)$ one
  needs to renormalize the metric $\kappa^{AB}$   \cite{Mohammedi:1993rg,FigueroaO'Farrill:1994hx,Figueroa-OFarrill:1996NuPhB.458..137F}. The relevant
  metric is $\Omega^{AB}=\kappa^{AB}+K^{AB}/2$ where
  $K^{AB}$ is the Killing form (see Section
  \ref{sc:FiniteTakiff}). The energy momentum tensor is then given as
  the normal ordered product
\begin{align}
  \label{eq:EnergyMomentum}
  T(z)=\frac{1}{2}\Omega_{AB}(J^AJ^B)(z)\;.
\end{align}
  The matrix $\Omega_{AB}$ is the inverse of the metric
  $\Omega^{AB}$. If this metric is not invertible the model is said to be at the critical level and the energy momentum tensor does not exists.\footnote{The latter case is still of interest since the center of the affine Lie algebra is significantly enlarged at the critical level, with important implications for the theory of integrable systems, see \cite{Feigin:1994CMaPh.166...27F}.}
  
  Since all these facts are well-established and frequently referred
  to in the literature we will not repeat the calculations (see however \cite{Rasmussen:2019NuPhB.94514680R}). Instead we
  focus on the derivation of the central charge for the special case when $\h=\fT_p(\g)$ is a Takiff superalgebra. The general
  expression for the central charge is given by   \cite{Mohammedi:1993rg,FigueroaO'Farrill:1994hx,Figueroa-OFarrill:1996NuPhB.458..137F}
\begin{align}
  \label{eq:CentralCharge}
  c=\Omega_{AB}\kappa^{AB}
    =\str\bigl(\Omega^{-1}\kappa\bigr)\;.
\end{align}
  Our analysis will show that in the case of Takiff superalgebras
  (with $p\geq2$), the central charge can be evaluated explicitly,
  thereby giving rise to the value
\begin{align}
  \label{eq:CentralChargeResult}
  c=p\,\sdim\g\ \ ,
\end{align}
  independently of the choice of metric. That the result is an integer is actually an immediate consequence of general statements that have been established in \cite{FigueroaO'Farrill:1994hx,Figueroa-OFarrill:1996NuPhB.458..137F}.

  In order to understand the result \eqref{eq:CentralChargeResult} we need to have more explicit expressions for the original metric $\kappa^{AB}$ and the renormalized metric $\Omega^{AB}$. Since the latter is obtained from the former by addition of the Killing form $K^{AB}$ we focus on $\kappa^{AB}$ first, returning to our original labels $a$ and $s$. According to Section \ref{sc:FiniteTakiff}, the metric $\kappa$ may be written in block-diagonal form as
\begin{align}
  \kappa
  =\mat\kappa_0&\kappa_1&\kappa_2&\cdots&\kappa_{p-2}&\kappa_{p-1}\\
     \kappa_1&\kappa_2&\kappa_3&\cdots&\kappa_{p-1}&0\\
     \kappa_2&\kappa_3&\kappa_4&\cdots&0&0\\
     \vdots&\vdots&\vdots&\ddots&\vdots&\vdots\\
     \kappa_{p-2}&\kappa_{p-1}&0&\cdots&0&0\\
     \kappa_{p-1}&0&0&\cdots&0&0\tam\;,
\end{align}
  where we used the abbreviations $\kappa_s^{ab}=\langle J^a,J^b\rangle_s$. The upper anti-block-triangular structure makes clear that this matrix possesses an inverse provided $\kappa_{p-1}$ is non-degenerate, confirming the findings of Section~\ref{sc:InvariantForms}. The precise form of the inverse $\kappa^{-1}$ is not relevant for the central charge. We only need to know that the inverse has a lower anti-block-triangular structure with $\kappa_{p-1}^{-1}$ on the anti-block-diagonal and vanishing entries in the upper left part.

  In the renormalized metric $\Omega$ which defines the energy momentum tensor, the contribution $\kappa_0$ is renormalized to a new metric $\kappa'_0$ (whose form is not important for the central charge), see Eq.~\eqref{eq:TakiffKilling}. Using the general formula \eqref{eq:CentralCharge}, the central charge of the  energy momentum tensor can finally be evaluated to be
\begin{align}
  c=\str\mat0&0&\kappa_{p-1}^{-1}\\0&\kappa_{p-1}^{-1}&\ast\\\kappa_{p-1}^{-1}&\ast&\ast\tam\mat\ast&\ast&\kappa_{p-1}\\\ast&\kappa_{p-1}&0\\\kappa_{p-1}&0&0\tam
  =\str\mat\idop&0&0\\\ast&\idop&0\\\ast&\ast&\idop\tam
  =p\sdim(\g)\;.
\end{align}
  The asterisk $\ast$ symbolizes entries which are known and which can be written down but which are not relevant for the final result. During the calculation it is important to note that the super-part of the supertrace is insensitive to the block structure.

\subsection{\label{sc:AffinizationTakiffization}Affinization commutes
  with Takiffization}

  Instead of starting with a finite dimensional Takiff superalgebra
  and considering its affinization one can also first affinize a
  finite dimensional Lie superalgebra and then Takiffize the
  result. We will now investigate in which sense both procedures can be
  regarded as equivalent.

  The main difference in both approaches is that the affinization
  $\ag$ of a Lie superalgebra $\g$ with invariant form $\langle\cdot,\cdot\rangle$ will introduce one central element
  $K$ which will then be duplicated upon Takiffization. The
  resulting elements $K_s=K\otimes\Theta^s$ are all central. On the other hand, if we first proceed to the Takiff superalgebra and then affinize there will be a single central element $K$. We will now explain this problem in more detail and point out how this apparent mismatch can be resolved on the level of representations by absorbing the choice of some free constants (``levels'') into the metric.
  
  If $J^a$ is a basis of generators of $\g$ then $J^a\otimes t^n$ together with the central element $K$ is a basis of generators of $\ag$. The commutation relations read
\begin{align}
  \bigl[J^a\otimes t^m,J^b\otimes t^n\bigr]
  =[J^a,J^b]\otimes t^{m+n}+\delta_{m+n,0}\,\langle J^a,J^b\rangle\,K\;.
\end{align}
  In a second step we Takiffize this algebra and end up with a basis of generators $J^a\otimes t^n\otimes\Theta^s$ and $K\otimes\Theta^s$ and commutation relations
\begin{align}
  \bigl[J^a\otimes t^m\otimes\Theta^r,J^b\otimes t^n\otimes\Theta^s\bigr]
  =[J^a,J^b]\otimes t^{m+n}\otimes\Theta^{r+s}+\delta_{m+n,0}\,\langle J^a,J^b\rangle\,K\otimes\Theta^{r+s}\;.
\end{align}
  We notice that there are $p$ distinct central elements $K_s=K\otimes\Theta^s$ appearing on the right hand side.
  
  On the other hand we can start with the natural basis $J^a\otimes\Theta^s$ of the Takiff superalgebra $\fT_p(\g)$ and the associated family of metrics $\langle\cdot,\cdot\rangle_s$. In the associated affinization the basis consists of $J^a\otimes\Theta^s\otimes t^n$ as well as a central element $K$ and the commutation relations become
\begin{align}
  \bigl[J^a\otimes\Theta^r\otimes t^m,J^b\otimes\Theta^s\otimes t^n\bigr]
  =[J^a,J^b]\otimes\Theta^{r+s}\otimes t^{m+n}
   +\delta_{m+n,0}\,\langle J^a,J^b\rangle_{r+s}\,K\;.
\end{align}
  So, while there is an obvious bijection between the generators $J^a\otimes t^m\otimes\Theta^r$ of the first and $J^a\otimes\Theta^r\otimes t^m$ of the second approach, the number of central elements is different and consequently there is no isomorphism between the two superalgebras in question.

  On the other hand the first approach is employing a unique metric $\langle\cdot,\cdot\rangle$ while the second approach makes use of a family of metrics $\langle\cdot,\cdot\rangle_s$ where $s=0,\ldots,p-1$. This permits to precisely match the number of free parameters if the central elements are treated as numbers, e.g.\ if we think about the action of the superalgebras in a suitable representation. In that case we could make the assumptions $K\otimes\Theta^s=k_s\in\Complex$ and $K=k\in\Complex$ together with the identification
\begin{align}
  k\,\langle\cdot,\cdot\rangle_s
  =k_s\,\langle\cdot,\cdot\rangle\;.
\end{align}
  Morally, the Takiffization thus commutes with affinization if the free parameters in both constructions are chosen appropriately.

\subsection{Automorphisms}

  In Section~\ref{sc:Automorphisms} we have established that any automorphism $\Omega$ of a Lie superalgebra $\g$ can be lifted to the associated Takiff superalgebra $\fT_p(\g)$. It is a matter of straightforward calculation that this automorphism also lifts to the affinization $\widehat{\fT}_p(\g)$ provided it is compatible with the family of metrics $\langle\cdot,\cdot\rangle_{s}$ used to construct the central extension. Indeed, with the definitions $\Omega(X\otimes\Theta^s\otimes t^m)=\Omega(X)\otimes\Theta^s\otimes t^m$ and $\Omega(K)=K$ we immediately find
\begin{align}
  &\hspace{-1.5cm}\bigl[\Omega(X\otimes\Theta^s\otimes t^m),\Omega(Y\otimes\Theta^r\otimes t^n)\bigr]\nonumber\\[2mm]
  &=\bigl[\Omega(X)\otimes\Theta^s\otimes t^m,\Omega(Y)\otimes\Theta^r\otimes t^n\bigr]\nonumber\\[2mm]
  &=\bigl[\Omega(X),\Omega(Y)\bigr]\otimes\Theta^{r+s}\otimes t^{m+n}
   +m\,\bigl\langle\Omega(X),\Omega(Y)\bigr\rangle_{r+s}\,\delta_{m+n,0}\,K\nonumber\\[2mm]
  &=\Omega\bigl([X,Y]\bigr)\otimes\Theta^{r+s}\otimes t^{m+n}
   +m\,\bigl\langle X,Y\bigr\rangle_{r+s}\,\delta_{m+n,0}\,K\nonumber\\[2mm]
  &=\Omega\bigl([X,Y]\otimes\Theta^{r+s}\otimes t^{m+n}\bigr)
    +m\,\bigl\langle X,Y\bigr\rangle_{r+s}\,\delta_{m+n,0}\,\Omega(K)\nonumber\\
  &=\Omega\bigl([X,Y]\otimes\Theta^{r+s}\otimes t^{m+n}
  +m\,\bigl\langle X,Y\bigr\rangle_{r+s}\,\delta_{m+n,0}\,K\bigr)\;.
\end{align}
  Moreover, every such isometric automorphism preserves the Sugawara energy momentum tensor defined in Eq.\ \eqref{eq:Sugawara} and hence gives rise to an automorphism of the full underlying vertex operator algebra.\footnote{In view of the necessary renormalization of the metric we used here that the Killing form is invariant under any automorphism.} This property is important for the discussion of conformal boundary conditions (D-branes) since the automorphisms may be used to glue chiral currents at the boundary of the world-sheet, see \cite{Behrend:1999bn} and references therein.

  It is conceivable that there are many more automorphisms of  $\widehat{\fT}_p(\g)$ that are relevant in a CFT context. To name just a single example, spectral flow automorphisms which involve shifts of the mode indices and which therefore do not have an analogue in the underlying finite dimensional Takiff super\-algebra $\fT_p(\g)$ play a significant role in WZW models based on non-compact groups~\cite{Maldacena:2000hw} or at fractional level~\cite{Gaberdiel:2001ny,Lesage:2002ch}. Since the discussion of spectral flow automorphisms will rely on additional structure on $\g$ and hence $\widehat{\fT}_p(\g)$ we will refrain from presenting further details in this note.


\section{\label{sc:Examples}Examples}

\subsection{Abelian Takiff superalgebras}

  As was mentioned already in \cite{Babichenko:2013JPhA...46l5204B}, the Takiff
  superalgebra $\fT_p(\g)$ associated with an abelian superalgebra
  $\g$ is abelian. For this reason, the whole structure coming with
  the natural $\Integer$-grading of a Takiffization is somewhat void
  since there is no intrinsic reason for additional structures such as
  metrics and automorphisms to be compatible with the
  $\Integer$-grading. As abelian subalgebras may be interesting from a
  physical point of view but not from a Takiff point of view, we will
  disregard them in what follows.

\subsection{Takiff superalgebras of order 2}

  Restricting our attention to Takiff superalgebras of order $p=2$
  allows us to make contact to the results of Babichenko and Ridout
  \cite{Babichenko:2013JPhA...46l5204B}. To simplify notation, we will identify
  $J^a\otimes1$ with $J_0^a$ and $J^a\otimes\Theta$ with
  $J_1^a$. According to the results of Section
  \ref{sc:FiniteTakiffDefinition}, the assignment
\begin{align}
  \langle J_0^a,J_0^b\rangle
  =\kappa_0(J^a,J^b)
  =\kappa_0^{ab}\ ,
  \qquad
  \langle J_0^a,J_1^b\rangle
  =\kappa_1(J^a,J^b)
  =\kappa_1^{ab}\ ,
  \qquad
  \langle J_1^a,J_1^b\rangle
  =0
\end{align}
  defines a natural invariant form on $\fT_2(\g)$. If $\g$ is simple,   we may write $\kappa_0=k_0\kappa$ and $\kappa_1=k_1\kappa$ where $\kappa$ is the suitably normalized standard metric on $\g$ and $k_0$ and $k_1$ are two numbers. In matrix form the metric and its inverse now assume
  the simple form
\begin{align}
  \kappa
  =\mat k_0\kappa^{ab}&k_1\kappa^{ab}\\k_1\kappa^{ab}&0\tam
  \qquad\text{ and }\qquad
  \kappa^{-1}
  =\frac{1}{k_1^2}\mat 0&k_1\kappa_{ab}\\k_1\kappa_{ab}&-k_0\kappa_{ab}\tam\;.
\end{align}
  Here $\kappa_{ab}$ denotes the inverse of $\kappa^{ab}$. In order to write down the energy momentum tensor for the affinization of $\fT_2(\g)$ we need to renormalize the level $k_0$ to $k_0+2g^\vee$ where $g^\vee$ is the dual Coxeter number of $\g$. We would like to stress that Eq.~\eqref{eq:TakiffKilling} implies a renormalization by $pg^\vee$ instead of the standard $g^\vee$ where $p=2$ in the present case. Comparing with the general formula \eqref{eq:EnergyMomentum}, we end up with
\begin{align}
  T(z)
  =\frac{1}{2k_1}\bigl[(J_0J_1)(z)+(J_1J_0)(z)\bigr]-\frac{k_0+2g^\vee}{2k_1^2}\,(J_1J_1)(z)\;.
\end{align}
  The result coincides with the expressions found in \cite{Babichenko:2013JPhA...46l5204B,Rasmussen:2017NuPhB.922..435R}. However, while in the latter references the definition of the energy momentum tensor seemed to be somewhat ad hoc we now understand its precise origin.

\subsection{Takiff superalgebras of order 3}

  In order to get some intuition for higher order Takiff
  superalgebras, let us finally have a closer look at the case
  $p=3$. In this case we work with generators
  $J_s^a=J^a\otimes\Theta^s$. Restricting our attention to simple Lie
  superalgebras $\g$, the metric and its inverse may be written as
\begin{align}
  \kappa
  =\mat k_0\kappa&k_1\kappa&k_2\kappa\\k_1\kappa&k_2\kappa&0\\k_2\kappa&0&0\tam
  \quad\text{ and }\quad
  \kappa^{-1}
  =\frac{1}{k_2^3}\mat 0&0&k_2^2\kappa^{-1}\\0&k_2^2\kappa^{-1}&-k_1k_2\kappa^{-1}\\k_2^2\kappa^{-1}&-k_1k_2\kappa^{-1}&(k_1^2-k_0k_2)\kappa^{-1}\tam\;.
\end{align}
  The corresponding energy momentum tensor reads, again taking into account the proper renormalization of the level $k_0$ by $pg^\vee=3g^\vee$,
\begin{align}
  T(z)
  &=\frac{1}{2k_2}\Bigl[(J_0J_2)(z)+(J_1J_1)(z)+(J_2J_0)(z)\Bigr]
       -\frac{k_1}{2k_2^2}\bigl[(J_1J_2)(z)+(J_2J_1)(z)\bigr]\nonumber\\[2mm]
  &\qquad\qquad+\frac{k_1^2-(k_0+3g^\vee)k_2}{2k_2^3}\,(J_2J_2)(z)\;.
\end{align}
  Of course the analysis could easily be extended to higher values of the order $p$. However, since the expressions get increasingly lengthy we refrain from presenting explicit formulas. The interested reader may refer to Ref.~\cite{Rasmussen:2019NuPhB.94514680R} where Takiff algebras are constructed and discussed from the perspective of Galilean contractions. In a classical context, i.e.\ without renormalization of the metric, these formulas have also been obtained in Ref.~\cite{Vicedo:2017arXiv170104856V}.

\subsubsection*{Acknowledgements}

  TQ would like to thank David Ridout for comments on the manuscript and the referees for their constructive feedback. This research was conducted by the Australian Research Council Centre of Excellence for Mathematical and Statistical Frontiers (project number CE140100049) and partially funded by the Australian Government.

\section{Conclusions}

  We have provided a detailed discussion of Takiff superalgebras
  $\fT_p(\g)$ and their relation to conformal field theory. We showed that every Takiff superalgebra equipped with a generic
  metric defines an associated WZW model with a canonical energy momentum tensor that is obtained by means of the standard Sugawara
  construction. As a by-product we established that, in this sense, the conformal field
  theories studied in \cite{Babichenko:2013JPhA...46l5204B} (for $p=2$) are
  ordinary WZW models. We also extended the analysis to higher order Takiff superalgebras with $p>2$ and this allowed us to connect to recent work on Galilean contractions of affine Lie algebras \cite{Rasmussen:2017NuPhB.922..435R,Rasmussen:2019NuPhB.94514680R}. The main benefit of our result is that it provides a natural geometric interpretation of these conformal field theories and hence access to a rich toolkit involving, e.g.\ methods of harmonic analysis. Also, many questions, e.g. of representation theoretic nature, that may be quite intricate to discuss directly on the level of infinite dimensional Lie algebras can presumably be reduced to the finite dimensional setting or at least informed by the latter.
  
  Our results are very general and with generality in mind we had to limit our exposition. In particular, we did not discuss the representation theory of the finite and affine Takiff superalgebras we constructed since we did not make any assumptions on the underlying Lie superalgebra $\g$ except for the existence of an invariant metric. For the actual solution of specific models this is the most urgent point that needs to be addressed. It is likely that key insights can be gained from induction of representations from $\g$ to $\fT_p(\g)$ or from lifting a potential root space decomposition from $\g$ to $\fT_p(\g)$ and analyzing the associated Verma modules. In light of the results of \cite{Rasmussen:2017NuPhB.922..435R,Rasmussen:2019NuPhB.94514680R} an alternative avenue may consist in studying the effect of performing contractions on direct sums of $\g$ or its affinizations $\ag$.
  This last perspective has been very successful when solving the $H_4$ model whose underlying Heisenberg group arises from a contraction of $SL(2,\Real)\times U(1)$ \cite{D'Appollonio:2003dr,D'Appollonio:2003ym,Bianchi:2004vf,D'Appollonio:2004pm}. All these considerations can and should be complemented by considering the harmonic analysis on Lie supergroups associated with $\fT_p(\g)$ and its relation to the harmonic analysis on Lie supergroups associated with $\g$.
  
  Recent work on WZW models based on simple Lie superalgebras has shown that the types of representations that play a role in the solution of the CFT differ quite drastically depending on whether the level is fractional or integral \cite{Lesage:2002ch,Creutzig:2012sd,Kawasetsu:2019CMaPh.368..627K,Kawasetsu:2019arXiv190602935K}. It is thus important to gain a better understanding of what it means for Takiff superalgebras to have an integral as opposed to a fractional level. This is also deeply related to the geometric question of how to describe integral 3-forms or, from a more elaborate perspective, bundle gerbes \cite{Gawedzki:2002RvMaP..14.1281G} on the associated Lie supergroup. Indeed, an integral 3-form is required in order to be able to define the WZW Lagrangian which includes a topological Wess-Zumino term. As far as we are aware, most of these questions have not been addressed systematically for non-reductive Lie groups, yet alone supergroups, and hence provide a strong motivation for further work in this direction. Let us also note that all of our considerations should admit a natural generalization to multi-graded Takiff superalgebras that have been introduced in the recent paper \cite{Ragoucy:2020arXiv200208637R}.
  
  Let us finally observe that Takiff superalgebras have played a prominent role recently beyond CFT, in the discussion of integrable systems \cite{Vicedo:2017arXiv170104856V,Delduc:2019JHEP...06..017D,Lacroix:2020JPhA...53y5203L}. We hope that the point of view developed in this paper will also prove useful in that connection.  


\begin{thebibliography}{10}

\bibitem{Goddard:1985vk}
P.~Goddard, A.~Kent and D.~I. Olive, {\it Virasoro algebras and coset space
  models},  {\rm Phys. Lett.} {\bf B152} (1985) 88.

\bibitem{Felder:1988CMaPh.117..127F}
G.~{Felder}, K.~{Gawedzki} and A.~{Kupiainen}, {\it {Spectra of
  Wess-Zumino-Witten models with arbitrary simple groups}},  {\rm Comm. Math.
  Phys.} {\bf 117} (1988) 127--158.

\bibitem{Bershadsky:1989mf}
M.~Bershadsky and H.~Ooguri, {\it Hidden {$sl(n)$} symmetry in conformal field
  theories},  {\rm Commun. Math. PHYS.} {\bf 126} (1989) 49.

\bibitem{Feigin:1990pn}
B.~Feigin and E.~Frenkel, {\it Quantization of the {Drinfeld-Sokolov}
  reduction},  {\rm Phys. Lett.} {\bf B246} (1990) 75--81.

\bibitem{Witten:1983ar}
E.~Witten, {\it Nonabelian bosonization in two dimensions},  {\rm Commun. Math.
  Phys.} {\bf 92} (1984) 455--472.

\bibitem{Knizhnik:1984nr}
V.~G. Knizhnik and A.~B. Zamolodchikov, {\it Current algebra and {Wess-Zumino}
  model in two dimensions},  {\rm Nucl. Phys.} {\bf B247} (1984) 83--103.

\bibitem{Gepner:1986wi}
D.~Gepner and E.~Witten, {\it String theory on group manifolds},  {\rm Nucl.
  Phys.} {\bf B278} (1986) 493.

\bibitem{Nappi:1993ie}
C.~R. Nappi and E.~Witten, {\it A {WZW} model based on a nonsemisimple group},
  {\rm Phys. Rev. Lett.} {\bf 71} (1993) 3751--3753
  [\href{http://arXiv.org/abs/hep-th/9310112}{{\tt hep-th/9310112}}].

\bibitem{FigueroaO'Farrill:1994hx}
J.~M. Figueroa-O'Farrill and S.~Stanciu, {\it Nonsemisimple {Sugawara}
  constructions},  {\rm Phys. Lett.} {\bf B327} (1994) 40--46
  [\href{http://arXiv.org/abs/hep-th/9402035}{{\tt hep-th/9402035}}].

\bibitem{Figueroa-OFarrill:1996NuPhB.458..137F}
J.~M. {Figueroa-O'Farrill} and S.~{Stanciu}, {\it {Nonreductive WZW models and
  their CFTs}},  {\rm Nuclear Physics B} {\bf 458} (1996), no.~1 137--164
  [\href{http://arXiv.org/abs/hep-th/9506151}{{\tt hep-th/9506151}}].

\bibitem{D'Appollonio:2003dr}
G.~D'Appollonio and E.~Kiritsis, {\it String interactions in gravitational wave
  backgrounds},  {\rm Nucl. Phys.} {\bf B674} (2003) 80--170
  [\href{http://arXiv.org/abs/hep-th/0305081}{{\tt hep-th/0305081}}].

\bibitem{D'Appollonio:2003ym}
G.~D'Appollonio, {\it Gravitational waves from {WZW} models},  {\rm Class.
  Quant. Grav.} {\bf 21} (2004) S1329--1336
  [\href{http://arXiv.org/abs/hep-th/0312206}{{\tt hep-th/0312206}}].

\bibitem{Bianchi:2004vf}
M.~Bianchi, G.~D'Appollonio, E.~Kiritsis and O.~Zapata, {\it String amplitudes
  in the {Hpp}-wave limit of {$AdS_3\times S^3$}},  {\rm JHEP} {\bf 04} (2004)
  074 [\href{http://arXiv.org/abs/hep-th/0402004}{{\tt hep-th/0402004}}].

\bibitem{D'Appollonio:2004pm}
G.~D'Appollonio and E.~Kiritsis, {\it D-branes and {BCFT} in {Hpp}-wave
  backgrounds},  {\rm Nucl. Phys.} {\bf B712} (2005) 433
  [\href{http://arXiv.org/abs/hep-th/0410269}{{\tt hep-th/0410269}}].

\bibitem{Babichenko:2013JPhA...46l5204B}
A.~{Babichenko} and D.~{Ridout}, {\it {Takiff superalgebras and conformal field
  theory}},  {\rm J. Phys.} {\bf A46} (2013), no.~12 125204
  [\href{http://arXiv.org/abs/1210.7094}{{\tt 1210.7094}}].

\bibitem{Rasmussen:2017NuPhB.922..435R}
J.~{Rasmussen} and C.~{Raymond}, {\it {Galilean contractions of W-algebras}},
  {\rm Nucl. Phys.} {\bf B922} (2017) 435--479
  [\href{http://arXiv.org/abs/1701.04437}{{\tt 1701.04437}}].

\bibitem{Rasmussen:2019NuPhB.94514680R}
J.~{Rasmussen} and C.~{Raymond}, {\it {Higher-order Galilean contractions}},
  {\rm Nucl. Phys.} {\bf B945} (2019) 114680
  [\href{http://arXiv.org/abs/1901.06069}{{\tt 1901.06069}}].

\bibitem{Babichenko:2015SIGMA..11..067B}
A.~{Babichenko} and T.~{Creutzig}, {\it Harmonic analysis and free field
  realization of the {Takiff} supergroup of {$GL(1|1)$}},  {\rm SIGMA} {\bf 11}
  (2015) 067 [\href{http://arXiv.org/abs/1411.1072}{{\tt 1411.1072}}].

\bibitem{Kac:1977em}
V.~G. Kac, {\it Lie superalgebras},  {\rm Adv. Math.} {\bf 26} (1977) 8--96.

\bibitem{Takiff:MR0281839}
S.~J. Takiff, {\it Rings of invariant polynomials for a class of {Lie}
  algebras},  {\rm Trans. Amer. Math. Soc.} {\bf 160} (1971) 249--262.

\bibitem{Schomerus:2005bf}
V.~Schomerus and H.~Saleur, {\it The {$GL(1|1)$} {WZW} model: {From}
  supergeometry to logarithmic {CFT}},  {\rm Nucl. Phys.} {\bf B734} (2006)
  221--245 [\href{http://arXiv.org/abs/hep-th/0510032}{{\tt hep-th/0510032}}].

\bibitem{Gotz:2006qp}
G.~{G{\"o}tz}, T.~Quella and V.~Schomerus, {\it The {WZNW} model on
  {$PSU(1,1|2)$}},  {\rm JHEP} {\bf 03} (2007) 003
  [\href{http://arXiv.org/abs/hep-th/0610070}{{\tt hep-th/0610070}}].

\bibitem{Saleur:2006tf}
H.~Saleur and V.~Schomerus, {\it On the {$SU(2|1)$} {WZW} model and its
  statistical mechanics applications},  {\rm Nucl. Phys.} {\bf B775} (2007)
  312--340 [\href{http://arXiv.org/abs/hep-th/0611147}{{\tt hep-th/0611147}}].

\bibitem{Quella:2007hr}
T.~Quella and V.~Schomerus, {\it Free fermion resolution of supergroup {WZNW}
  models},  {\rm JHEP} {\bf 09} (2007) 085
  [\href{http://arXiv.org/abs/0706.0744}{{\tt 0706.0744}}].

\bibitem{Quella:2013JPhA...46W4010Q}
T.~{Quella} and V.~{Schomerus}, {\it {Superspace conformal field theory}},
  {\rm J. Phys.} {\bf A46} (2013) 494010
  [\href{http://arXiv.org/abs/1307.7724}{{\tt 1307.7724}}].

\bibitem{Mohammedi:1993rg}
N.~Mohammedi, {\it On bosonic and supersymmetric current algebras for
  nonsemisimple groups},  {\rm Phys. Lett.} {\bf B325} (1994) 371--376
  [\href{http://arXiv.org/abs/hep-th/9312182}{{\tt hep-th/9312182}}].

\bibitem{Feigin:1994CMaPh.166...27F}
B.~{Feigin}, E.~{Frenkel} and N.~{Reshetikhin}, {\it {Gaudin model, Bethe
  Ansatz and critical level}},  {\rm Comm. Math. Phys.} {\bf 166} (1994), no.~1
  27--62 [\href{http://arXiv.org/abs/hep-th/9402022}{{\tt hep-th/9402022}}].

\bibitem{Behrend:1999bn}
R.~E. Behrend, P.~A. Pearce, V.~B. Petkova and J.-B. Zuber, {\it Boundary
  conditions in rational conformal field theories},  {\rm Nucl. Phys.} {\bf
  B570} (2000) 525--589 [\href{http://arXiv.org/abs/hep-th/9908036}{{\tt
  hep-th/9908036}}].

\bibitem{Maldacena:2000hw}
J.~M. Maldacena and H.~Ooguri, {\it Strings in {$AdS_3$} and the {$SL(2,R)$}
  {WZW} model. {I}},  {\rm J. Math. Phys.} {\bf 42} (2001) 2929--2960
  [\href{http://arXiv.org/abs/hep-th/0001053}{{\tt hep-th/0001053}}].

\bibitem{Gaberdiel:2001ny}
M.~R. Gaberdiel, {\it Fusion rules and logarithmic representations of a {WZW}
  model at fractional level},  {\rm Nucl. Phys.} {\bf B618} (2001) 407--436
  [\href{http://arXiv.org/abs/hep-th/0105046}{{\tt hep-th/0105046}}].

\bibitem{Lesage:2002ch}
F.~Lesage, P.~Mathieu, J.~Rasmussen and H.~Saleur, {\it The
  {$\widehat{su}(2)_{-1/2}$} {WZW} model and the {$\beta\gamma$} system},  {\rm
  Nucl. Phys.} {\bf B647} (2002) 363--403
  [\href{http://arXiv.org/abs/hep-th/0207201}{{\tt hep-th/0207201}}].

\bibitem{Vicedo:2017arXiv170104856V}
B.~{Vicedo}, {\it {On integrable field theories as dihedral affine Gaudin
  models}},  {\rm Int. Math. Res. Not.} (2018) arXiv:1701.04856
  [\href{http://arXiv.org/abs/1701.04856}{{\tt 1701.04856}}].

\bibitem{Creutzig:2012sd}
T.~Creutzig and D.~Ridout, {\it Modular data and {Verlinde} formulae for
  fractional level {WZW} models {I}},  {\rm Nucl. Phys.} {\bf B865} (2012)
  83--114 [\href{http://arXiv.org/abs/1205.6513}{{\tt 1205.6513}}].

\bibitem{Kawasetsu:2019CMaPh.368..627K}
K.~{Kawasetsu} and D.~{Ridout}, {\it {Relaxed highest-weight modules I: Rank 1
  cases}},  {\rm Comm. Math. Phys.} {\bf 368} (2019), no.~2 627--663
  [\href{http://arXiv.org/abs/1803.01989}{{\tt 1803.01989}}].

\bibitem{Kawasetsu:2019arXiv190602935K}
K.~{Kawasetsu} and D.~{Ridout}, {\it Relaxed highest-weight modules {II}:
  classifications for affine vertex algebras},  {\rm arXiv e-prints} (2019)
  arXiv:1906.02935 [\href{http://arXiv.org/abs/1906.02935}{{\tt 1906.02935}}].

\bibitem{Gawedzki:2002RvMaP..14.1281G}
K.~{Gaw{\c{e}}dzki} and N.~{Reis}, {\it {WZW Branes and Gerbes}},  {\rm Rev.
  Math. Phys.} {\bf 14} (2002), no.~12 1281--1334
  [\href{http://arXiv.org/abs/hep-th/0205233}{{\tt hep-th/0205233}}].

\bibitem{Ragoucy:2020arXiv200208637R}
E.~{Ragoucy}, J.~{Rasmussen} and C.~{Raymond}, {\it {Multi-graded Galilean
  conformal algebras}},  {\rm arXiv e-prints} (2020) arXiv:2002.08637
  [\href{http://arXiv.org/abs/2002.08637}{{\tt 2002.08637}}].

\bibitem{Delduc:2019JHEP...06..017D}
F.~{Delduc}, S.~{Lacroix}, M.~{Magro} and B.~{Vicedo}, {\it {Assembling
  integrable {\ensuremath{\sigma}}-models as affine Gaudin models}},  {\rm
  JHEP} {\bf 2019} (2019), no.~6 17
  [\href{http://arXiv.org/abs/1903.00368}{{\tt 1903.00368}}].

\bibitem{Lacroix:2020JPhA...53y5203L}
S.~{Lacroix}, {\it {Constrained affine Gaudin models and diagonal Yang-Baxter
  deformations}},  {\rm J. Phys.} {\bf A53} (2020), no.~25 255203
  [\href{http://arXiv.org/abs/1907.04836}{{\tt 1907.04836}}].

\end{thebibliography}

\def\cprime{$'$}
\providecommand{\href}[2]{#2}\begingroup\raggedright\endgroup

\end{document}